\documentclass{article}
\usepackage{epsfig}

\newcommand{\tx}[1]{\textrm{#1}}
\newcommand{\no}{\noindent}

\begin{document}

\begin{center}

\Large{\boldmath \textbf{The $\eta$ $\to$ $\pi^0\pi^0\gamma\gamma$ decay in Generalized $\chi$PT}} \\
\vspace{0.5cm}
\normalsize Mari\'{a}n Koles\'{a}r, Ji\v{r}\'{\i} Novotn\'{y}\\
\vspace{0.25cm}
\em{Charles University, Faculty of Mathematics and Physics,
 V Hole\v{s}ovi\v{c}k\'{a}ch 2, 180\,00~Praha~8, Czech Republic}
\end{center}

\begin{abstract}
 Calculations of $\eta$ $\to$ $\pi^0\pi^0\gamma\gamma$ decay in Generalized
 chiral perturbation theory are presented. Tree level and next-to-leading
 corrections are involved. Sensitivity to violation of the Standard
 counting is discussed.
\end{abstract}

 \section{Introduction}

 The $\eta(p\,) \to \pi^0(p_1)\,\pi^0(p_2)\,\gamma(k)\,\gamma(k')$ process
 is a rare decay, which has been recently studied by several authors in
 context of Standard chiral perturbation theory (S$\chi$PT), namely at the
 lowest order by Kn\"ochlein, Scherer and Drechsel \cite{Drechsel} and to
 next-to-leading by Bellucci and Isidori \cite{Belucci} and Ametller
 et al. \cite{Ametller}. The experimental interest for such a process comes
 from the anticipation of large number of $\eta$'s to be produced at various
 facilities.\footnote{according to \cite{Belucci II}, at DA$\Phi$NE about $10^8$
 decays per year} The goal of our computations is to add the result for
 the next-to-leading order in Generalized chiral perturbation theory (G$\chi$PT).
 The motivation is that one of the important contributions involve the
 $\eta\,\pi \to \eta\,\pi$ off-shell vertex which is very sensitive to the
 violation of the Standard scheme and thus this decay provides a possibility
 of its eventual observation. We have completed the calculations at the tree
 level, added 1PI one loop corrections, corrections to the
 $\eta\,\pi \to \eta\,\pi$ vertex and phenomenological corrections to the
 resonant contribution. These preliminary results we would like to present in this paper.

 \section{Kinematics and parameters}

\vspace{0.25cm}
 The amplitude of the process can be defined

\vspace{-0.25cm}
\begin{equation}
 \langle \pi ^0(p_1)\pi ^0(p_2)\gamma (k,\epsilon )\gamma (k^{^{\prime
}},\epsilon ^{^{\prime }})_{\rm out}|\eta (p)_{\rm in}\rangle =i(2\pi )^4\delta
^{(4)}(P_f-p){\cal M}_{fi}.
\end{equation}

 \no In the square of the amplitude summed over the polarizations $
 \overline{|{\cal M}_{fi}|^2}=\sum_{\rm pol.}|{\cal M}_{fi}|^2
 $ we integrated out all of the independent Lorentz invariants except the
 diphoton energy square

\begin{equation}
  s_{\gamma\gamma} = (\:k+k')^2 ,\quad
  0 <\, s_{\gamma\gamma} \leq\, (M_{\eta}-2M_{\pi})^2.
\end{equation}

 \no Our goal is to calculate the partial decay width $\tx{d}\Gamma$
 of the $\eta$ particle as the function of the diphoton energy square $s_{\gamma\gamma}$.

 At the lowest order, the S$\chi$PT does not depend on any unknown free order
 parameters. In contrast, there are two free parameters controlling the
 violation of the Standard picture in the Generalized scheme. We have
 chosen them as
\begin{equation}
  r\ =\ \frac{m_s}{\hat{m}}\,,\quad X_{GOR}\ =\ \frac{2B\hat{m}}{M_{\pi}^2}
\end{equation}

\no and their ranges are $r \sim r_1 - r_2 \sim 6 - 26\, ,\ 0\ \leq\ X_{\rm GOR}\
 \leq\ 1$. We use abbreviations for $\hat{m}=(m_u+m_d)/2$, $r_1=2 M_K/M_{\pi}-1$
 and $r_2=2 M_K^2/M_{\pi}^2-1$. The Standard values of these parameters are
 $r=r_2$ and $X_{\rm GOR}=1$. 

 \section{Tree level}

 At the $O(p^4)$ tree level, the amplitude has two contributions, with
 a pion and an eta propagator. The first one is resonant, `$\pi^0$-pole',
 the other is not, `$\eta$-tail'.

\begin{figure}[h]
\epsfig{figure=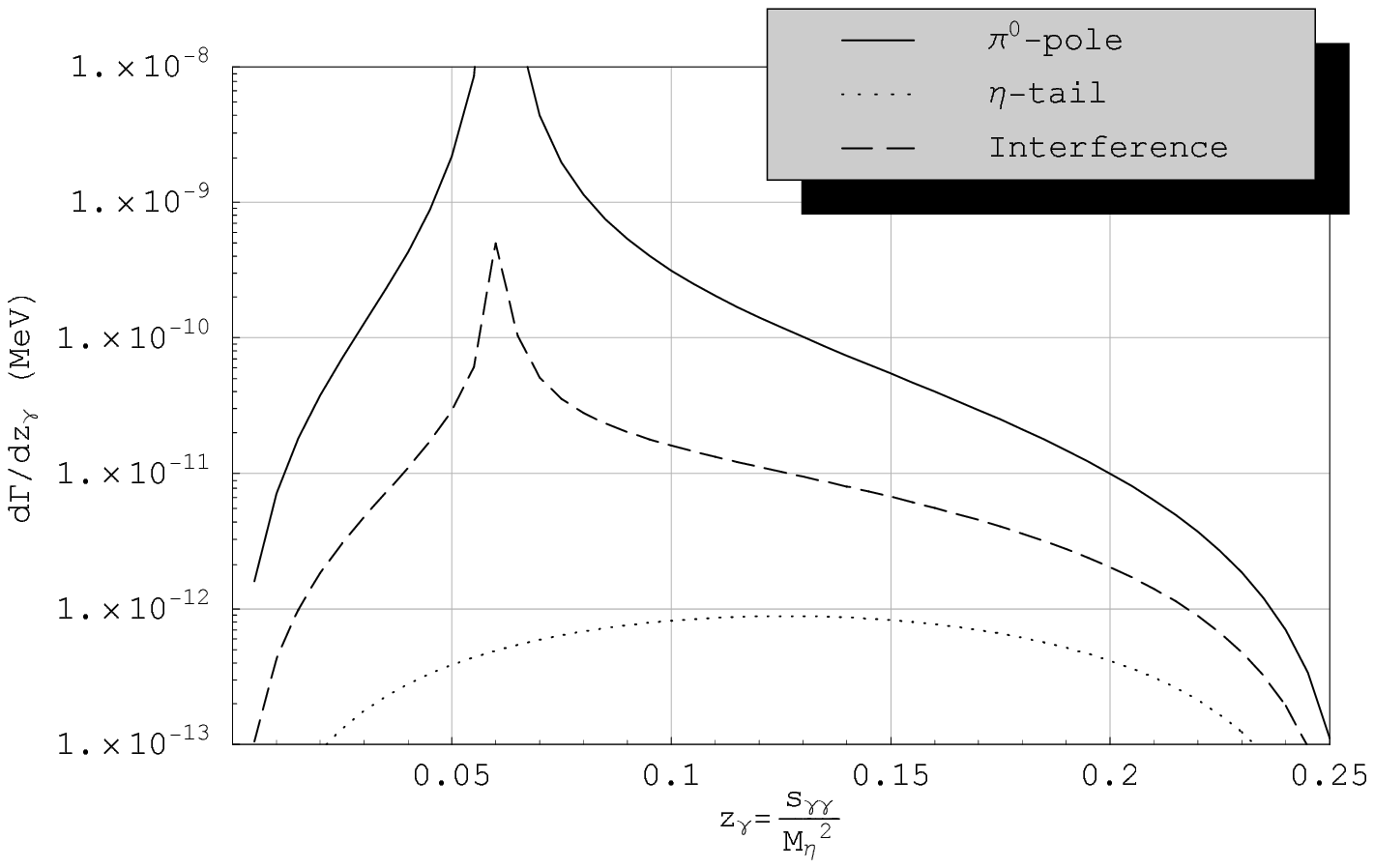,width=0.48\textwidth} 
 \hfill 
\epsfig{figure=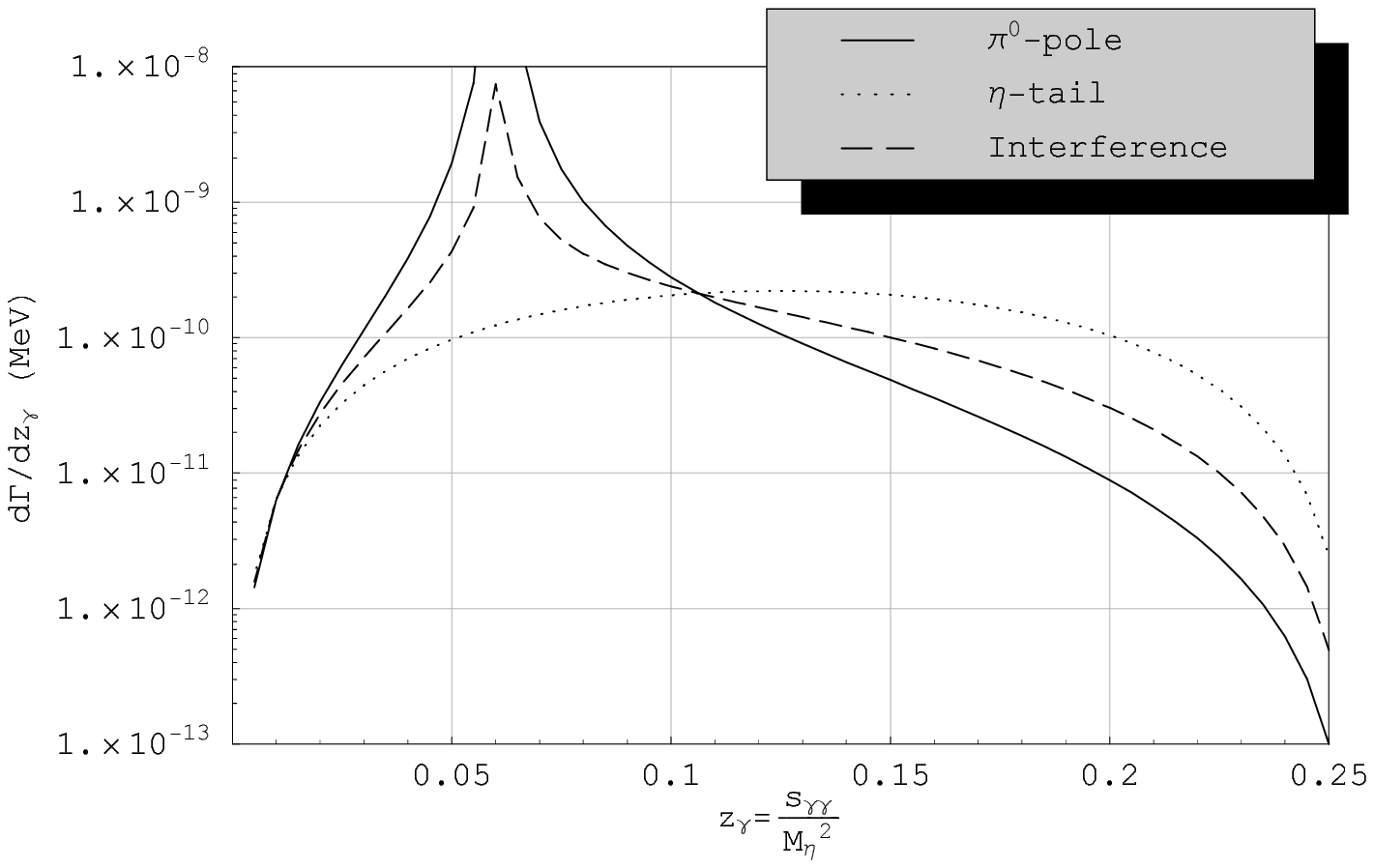,width=0.48\textwidth}
\caption{S$\chi$PT and G$\chi$PT tree level contributions to the partial
 decay rate $\tx{d}\Gamma/\tx{d}z_{\gamma}$}
\label{graph4}
\end{figure}

 The Standard values of the contributions to the partial decay rate and
 the maximum possible violation of the Standard counting ($r=r_1,X_{\rm GOR}=0$)
 are represented in Fig.~\ref{graph4}. The pole of the resonant contribution at
 $s_{\gamma\gamma}=M_{\pi}^2\,\sim\,0.06M_{\eta}^2$ is transparent. While
 in the Standard case it is fully dominant, in the Generalized scheme the
 $\eta$-tail could be determining in the whole area
 $s_{\gamma\gamma}>0.11M_{\eta}^2$. The reason can be found in the
 $\eta\,\pi \to \eta\,\pi$ vertex. Its contribution in the Generalized
 amplitude can jump up to 16 times its Standard value.

\begin{figure}[h]
 \begin{center}
\epsfig{figure=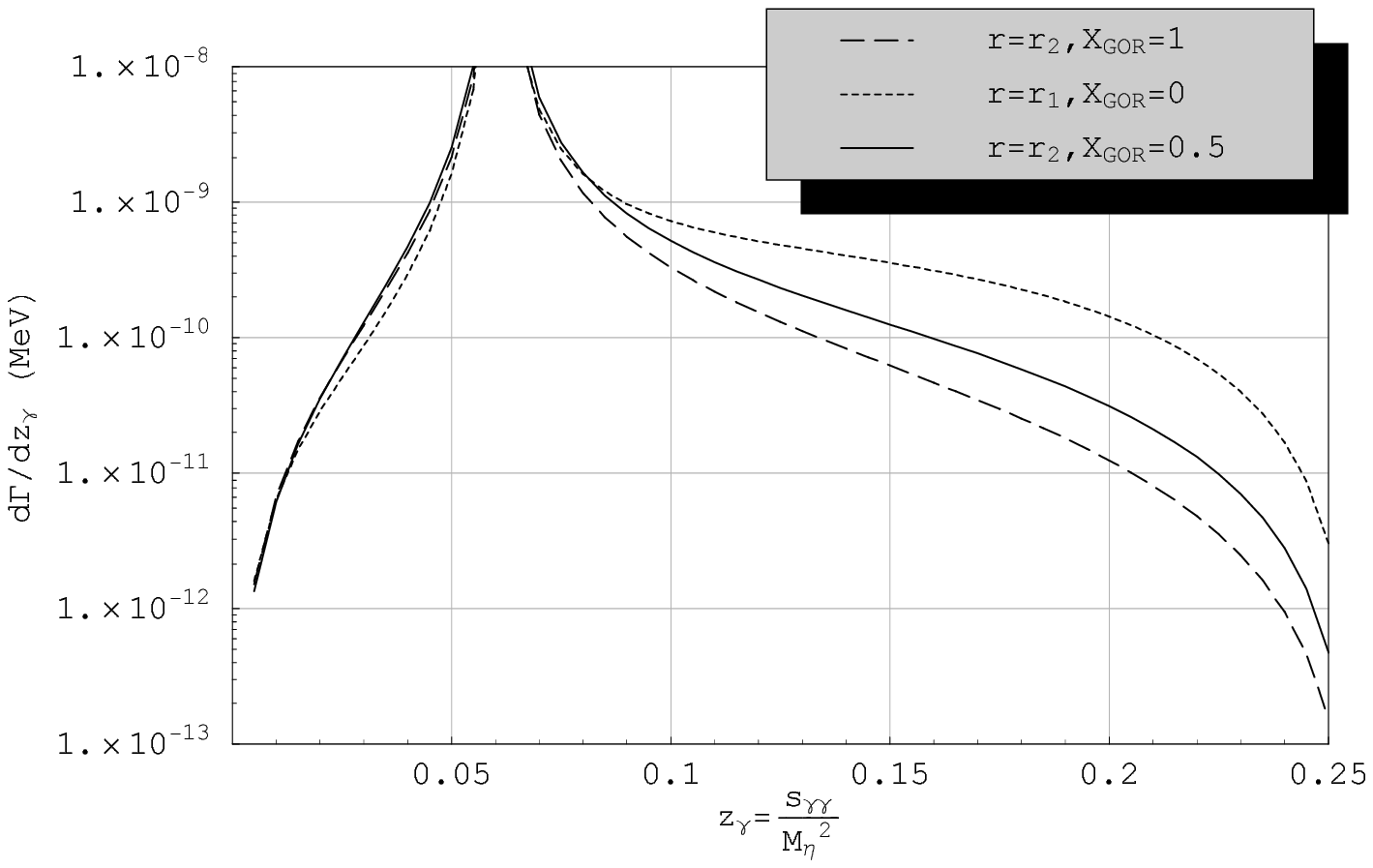,width=0.7\textwidth}
\caption{Full tree level decay width depending on the parameters $r$ and $X_{\rm GOR}$}
\label{graph6}
\vspace{-0.25cm}
 \end{center}\end{figure}

 The full decay width for the Standard ($r=r_2,X_{\rm GOR}=1$) and
 Generalized case ($r=r_2$,$X_{\rm GOR}=0.5$ and $r=r_1$,$X_{\rm GOR}=0$)
 is displayed in Fig.~\ref{graph6}. It can be seen, that even in the
 conservative intermediate case the change is quite interesting. 
 
 \section{One loop corrections}

 There are four distinct contributions at the next-to-leading order:
 one loop corrections to the $\pi^o$-pole and the $\eta$-tail, one
 particle irreducible diagrams (1PI) and counterterms.

 In the latter case we rely upon the results of \cite{Ametller}.
 Their estimate from vector meson dominated counterterms indicates,
 that it causes only a slight decrease of the full decay width.
 Because the estimate is the same for both schemes, for our purpose
 of studying the differences between them we can leave it for later investigation.

 More important are the corrections to the $\eta$-tail diagram.
 We did take into account the corrections to the $\eta\,\pi \to \eta\,\pi$
 vertex. These involve loop corrections and counterterms with many
 unknown higher order parameters. As a first approximation, we set
 these parameters equal to zero and estimated their effect through
 the remaining dependence on the renormalization scale. The scale
 was moved in the range from the mass of the $\eta$ to the mass of $\rho$-meson.

 We decided, similarly to \cite{Belucci}, to correct the $\pi^0$-pole
 amplitude by a phenomenological parametrization of the $\eta \to 3\pi^0$
 vertex and fix the parameters from experimental $\eta \to 3\pi^0$
 data. We made an estimate of its phase by expanding the $\eta \to 3\pi^0$
 one loop amplitude around the center of the Dalitz Plot.

\no In the 1PI amplitude, we neglected the suppressed kaon loops.

\begin{figure}[t]
\epsfig{figure=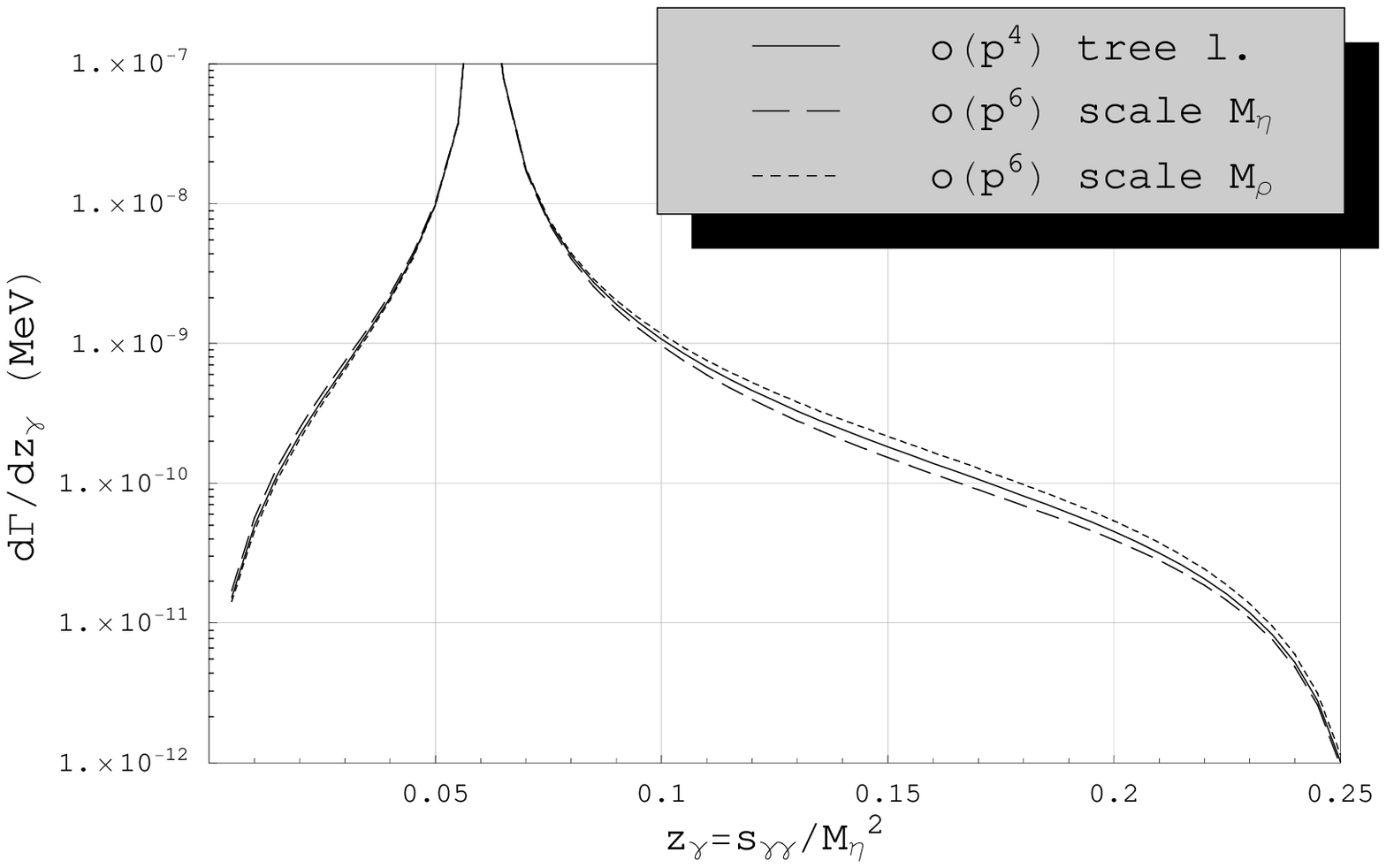,width=0.48\textwidth}
 \hfill 
\epsfig{figure=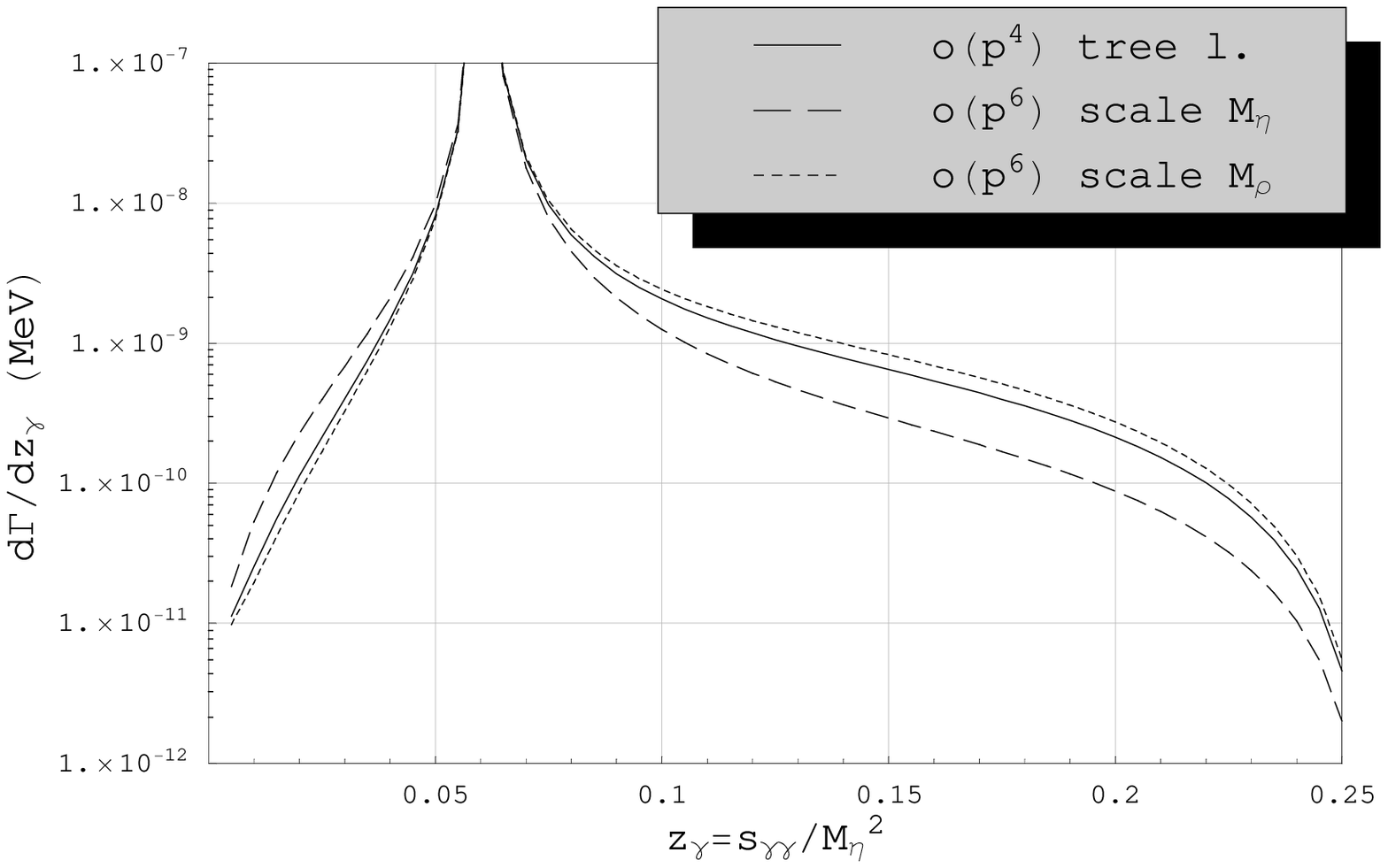,width=0.48\textwidth}
\label{graph1}
\vspace{0.25cm}
\epsfig{figure=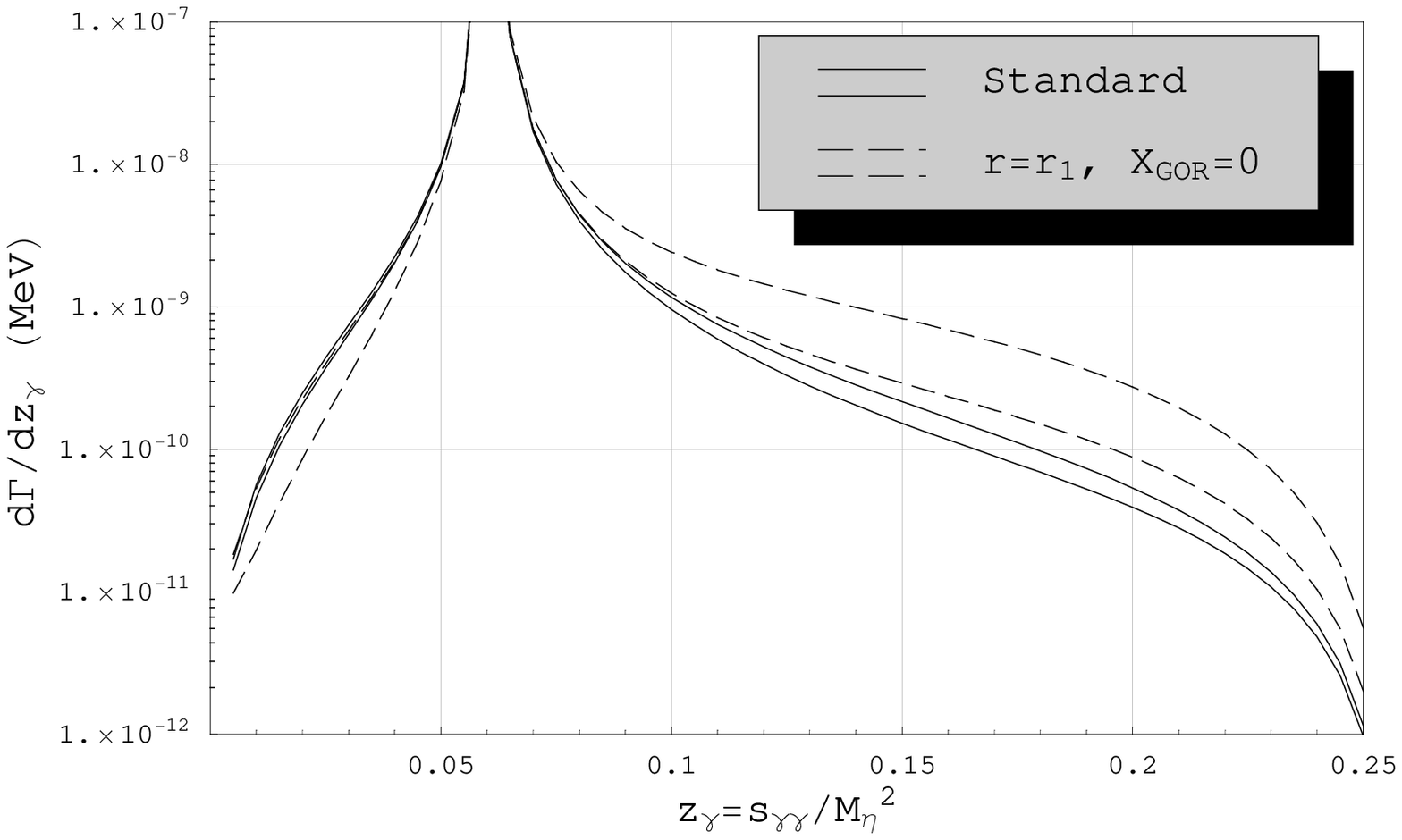,width=0.48\textwidth}
 \hfill 
\epsfig{figure=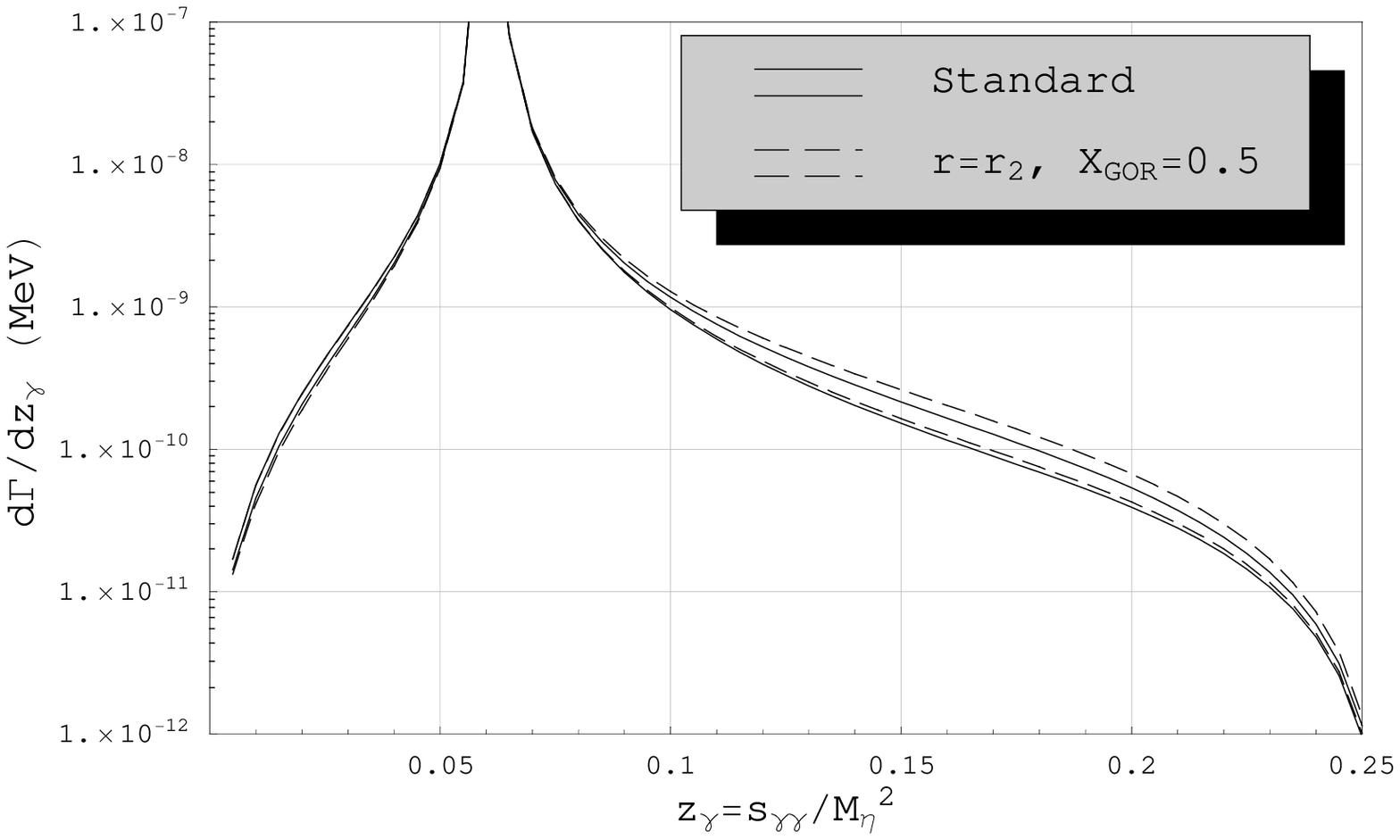,width=0.48\textwidth}
\caption{S$\chi$PT and G$\chi$PT tree level and one loop corrected full decay widths}
\label{graph2}
\end{figure}

 Fig.~\ref{graph2} represents the one loop corrected decay widths
 for the Standard and the maximum violation of the Standard scheme.
 The dependence on the renormalization scale is used to estimate the
 uncertainty in the unknown higher order coupling constants. We can
 see that the scale dependence is small in the Standard counting and
 not too terrible in the Generalized variant. In the case of the
 maximum violation of S$\chi$PT, the difference is big enough to not
 to be washed out by the uncertainty. However, in the conservative
 case $r=r_2$,$X_{\rm GOR}=0.5$ this is not true and the promising
 results from the tree level are lost.

\section{Conclusion}

 We have analyzed the $\eta\to\pi^0\pi^0\gamma\gamma$ decay to the
 next-to-leading order of chiral perturbation theory in its both variants.
 The tree level results are promising, the sensitivity to the change
 in parameters controlling the violation of the Stndard $\chi$PT is considerable.

 At the one loop level, we tried to estimate the uncertainty in the
 higher order couplings constants in the crucial $\eta\,\pi \to \eta\,\pi$
 vertex through their dependence on the renormalization scale. Although
 for big violation of the Standard case the difference is preserved,
 for the more realistic conservative case the output is not satisfactory.
 We would like to stress that these results are preliminary and there
 are several ways how to deal with the unknown order parameters. One
 of them is to take into account the vector mesons, similarly to the
 counterterm estimate in \cite{Ametller}. Other way is to treat the
 whole $\chi$PT expansion differently, with more caution, as developed
 in \cite{Stern}. This approach, called `resumed' $\chi$PT could provide
 results similar to the tree level case even if the one loop corrections are involved.

 \bigskip

 {\small This work was supported by program `Research Centers'
 (project number LN00A006) of the Ministry of Education of Czech Republic.}

\end{document}